# Familiarization tours for first-time users of highly automated cars: Comparing the effects of virtual environments with different levels of interaction fidelity[1]

*Mahdi Ebnali [1], Richard Lamb [2], Razieh Fathi [3]*
1. Industrial Engineering Department, University at Buffalo
2. Neurocognition Science Laboratory, University at Buffalo
3. Department of Computer Sciences, Rochester Institute of Technology

## Abstract

Research in aviation and driving has highlighted the importance of training as an effective approach to reduce the costs associated with the supervisory role of the human in automated systems. However, only a few studies have investigated the effect of pre-trip familiarization tours on highly automated driving. In the present study, a driving simulator experiment compared the effectiveness of four familiarization groups – control, video, low-fidelity virtual reality (VR), and high-fidelity VR – on automation trust and driving performance in several critical and non-critical transition tasks. The results revealed the positive impact of familiarization tours on trust, takeover, and handback performance at the first time of measurement. Takeover quality only improved when practice was presented in high-fidelity VR. After three times of exposure to transition requests, trust and transition performance of all groups converged to those of the high-fidelity VR group, demonstrating that: a) experiencing automation failures during the training may reduce costs associated with first failures in highly automated driving; b) the VR tour with high level of interaction fidelity is superior to other types of familiarization tour; and c) uneducated and less-educated drivers learn about automation by experiencing it. Knowledge resulting from this research could help develop cost-effective familiarization tours for highly automated vehicles in dealerships and car rental centers.

## 1. Introduction

*1.1 Automated driving: opportunities and challenges*

Advanced driver assistant systems (ADASs) have been introduced to the market to increase road transportation safety, improve the quality of travel, and optimize fuel efficiency. Autonomous vehicles are also expected to relieve traffic congestion and help to optimize spacing and density of driveways[1].

---

These technologies, which had been optional and deluxe features in the past, are included as standard equipment in today's cars. In the near future, highly automated vehicles, which will take full control of the car in some driving situations (e.g., designated highways), are also expected to be made available on the market. This technology provides longitudinal and lateral control in certain driving conditions and environments. It allows drivers to switch from an active controlling role to that of a supervisory controller, such that they do not have to control and monitor the road and can engage in non-driving-related tasks (NDRTs). However, due to technology limitations, infrastructure boundaries, and the dynamic nature of driving, the driver is called to engage in driving tasks and take manual control when the system surpasses the technical limits or when a takeover is scheduled [2]. Transitions between different modes are considered to be challenging tasks in automated driving and are associated with several human factor issues, such as erratic workload [3], complacency [4], and loss of situational awareness [5], leading to unsafe interaction and impaired takeover performance [6].

Although deploying appropriate automation logics and designing user-centered HMIs could facilitate drivers' interaction with automated systems, inappropriate and inaccurate mental models of the system have been considered underlying reasons for most of the aforementioned challenges. Studies have warned that uneducated drivers and those who have not been exposed to the limitations of assistive systems on the road tend to make false assumptions about the systems' functionality [7-9]. For example, a lack of understanding of the system's limitations may increase the driver's expectations of the automation functionality and result in overtrust and complacency. On the other hand, insufficient mental models of the system or encountering unexpected situations can lead drivers to underestimate the system's capabilities and discourage them from reaping the benefits of automated driving (distrust) [7, 9].

*1.2 Importance of training*

Several years of experience and practices in human factors have highlighted the importance of users' education as a necessary step in their successful and efficient cooperation with automated systems [6, 10-13]. Direct and indirect effects of training have been reflected in the reduction of catastrophic failures and incidents and improvement in the quality of human-automation interaction [14]. Considering these facts, car manufacturers are required to provide additional information in the owner's manual outlining how to operate automated features [15, 16]. Available manuals, however, suffer from several limitations which may reduce the quality of knowledge acquisition. First, they are mostly limited to basic operational information and do not cover the full range of scenarios a drive may encounter on the roads. Second, according to skill acquisition theories [17, 18], providing only rules as text-based instructions may result in limited learning outcomes, which are focused on declarative knowledge acquisition and do not encompass higher-order cognitive and procedural skills [19-21]. Third, while manuals are easy-to-access references, they are lengthy, and it is less likely that owners will read them thoroughly [9, 22]. Finally, compared to learning in the context, details memorized through texts are more likely to be misinterpreted, misremembered, or forgotten.

These limitations have motivated experts in human factors to explore other approaches, such as training drivers via simulators or in driving schools, where they have more chances to interact with automated systems and practice several possible scenarios to develop knowledge and procedural skills as well [23-25]. Although the findings are encouraging, these approaches suffer from major limitations; for example, it takes time to implement these types of training programs, and they require costly equipment or professional instructors, which may not be available to individuals.

Educating owners in dealership centers or car rental offices is a potential training method that is expected to increase drivers' knowledge of automated systems and could improve safety for first-time users [11, 26-28]. Currently, most automotive dealerships offer training or pre-trip demos before or after delivering cars to their owners. While interviews with owners revealed that drivers prefer to learn about their new cars in dealership centers, a recent study cast doubt on the quality and potency of such an approach. [29], in an investigation of how dealerships provide training and information regarding ADASs, reported that the effectiveness of training at the point of sale greatly varied depending on who delivered the training and what and how information was presented. Poorly delivered information (due to delivery methods, untrained salespersons, and low-quality explanations) and lack of standard materials are two main causes that significantly reduce the quality of learning through this approach and keep owners underinformed or misinformed about automated vehicles. The present study aims to bridge these gaps and explore the possible ways to improve the quality of pre-trip education programs for first-time users of highly automated cars.

The recent improvement in the quality of visual rendering and interaction fidelity, as well as the affordable expense of head-mounted display (HMD) VR, has encouraged researchers and companies to employ this technology for widespread use in training and education [30-34]. Despite the growing number of VR applications being developed for training in various domains, little research has been conducted to explore the effects of such technology in automated driving training. [23, 35], for example, used an HMD VR to examine its application in the acquisition of skills required for automated driving. Their results highlight the effectiveness of this training approach on reaction time (transition time from automated mode to manual mode) and users' self-reported measures (e.g. trust, acceptance, ease of use, understanding, and pleasantness). The results are encouraging, but they are scant, and the researchers did not employ a well-defined training protocol to cover essential educational requirements. Considering the availability of VR as a low-cost learning tool, further exploration and research are necessary to understand the affordances and efficacy of HMD VR as an immersive medium to prepare owners for automated driving upon the delivery of the car. Moreover, several studies have documented the benefits of interactive practices for training quality and educational outcomes. The critical question is; how does the interaction fidelity of a VR tour impacts first-time users' learning experiences and outcomes related to highly automated driving? This study aims to address the outstanding challenge of developing drivers' training paradigms by contrasting possible training approaches (video, VR with low-fidelity interaction, and VR with high-fidelity interaction) for highly automated driving.

*1.3 What do first-time drivers need to learn about automation?*

Although the available literature highlights the importance of drivers' familiarity with automation, little is known about content and best practices that effectively prepare the driver for highly automated driving. To select appropriate content and authentic tasks as the precursor and backbones of a training protocol, it is first important to understand what first-time users need to learn before driving a highly automated car. Validity and relevancy of the materials to the transfer task are highly correlated with desirable transfer, in which the responses learners have are consistent from the training to the real-world environment[36-38].

Based on lessons learned from the aviation industry, [11] outlined a set of knowledge standards for drivers of partially automated cars. *Knowledge about the automation* (information about functions, engineering logic, limitations, and equipment), *drivers* (maintaining awareness and responding to abnormal events), and *new driving tasks* are three essential kinds of knowledge that drivers should possess before they operate an automated car. An experimental study in partially automated driving [39], also covered parts of these educational requirements in an owner's manual to explore potential impacts of the training on automated driving safety. Their curriculums included knowledge about automated systems, physical equipment, working range of detection equipment, and requirements for optimal performance. Considering the minimum driver's training standards outlined for partially automated cars by [11], we extended each set of knowledge for highly automated driving (Figure 1).

Beyond the essential knowledge required of drivers regarding the features of an automated system, they also need to learn about and experience the tasks required for automated driving. A growing body of literature in human factors has identified transitions between different automation modes as the most challenging tasks the driver needs to perform safely [40, 41]. This concern is particularly more complicated for highly automated driving, where there is a wider avenue for problems associated with out-of-the-loop (OOTL) performance. This makes transitions to and from automated modes even more demanding, resulting in higher situation awareness (SA) loss and misunderstanding of what corrective actions need to be taken.

According to Gold, Naujoks [42] and the suggestion of existing evidence, we considered covering three types of tasks in the training protocol: *1) Critical takeover tasks (critical):* These takeover tasks constitute the most crucial part of the challenges associated with highly automated driving, and it is necessary to prepare the driver for such circumstances before the initial exposure to real driving. *2) Non-critical takeover tasks (non-critical)*: Non-critical takeover tasks have also been discussed as important activities in automated driving [43, 44]. Compared to critical takeover events, these scenarios have less criticality and urgency or are predictable, though familiarizing the driver with these kinds of non-critical system-initiated transitions could be essential to a pre-trip training program. *3) Handback tasks (non-critical)*: Some recent studies have also noted the importance of human-initiated transitions as everyday tasks in which the driver engages in automated driving [45]. In these cases, top-down processing is triggered based on an existing schema and prior information, and the driver, at a tactical level, evaluates the situation, anticipates, and finally relinquishes control to the automated system. There is still a limited understanding of drivers' behavior in these types of transitions, as well as how prior familiarization influences their performance.

Covering non-critical transition tasks in a familiarization tour has two potential benefits. First, non-critical transitions are common tasks and comprise the majority of interaction with an automated car. Second, learning how to appropriately use the automation in normal situations, for example, understanding how and when to relinquish control to the automation, helps drivers build richer mental models of the automation functionality.

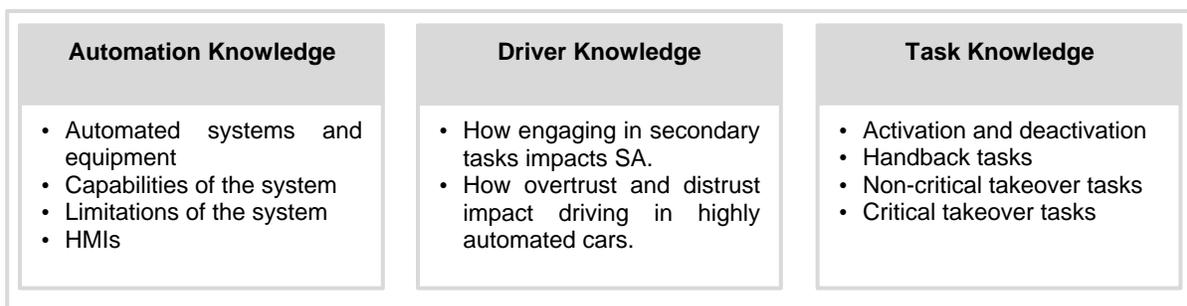

*Figure 1*. Materials covered in the familiarization tours, based on the minimum training standards proposed by [11] and regulations proposed for automated driving systems by [16]

According to the aforementioned training requirements (Figure 1), we designed three different types of familiarization tours: video, low-fidelity VR, and high-fidelity VR. All the tours shared the same content (automation knowledge, driver knowledge, and task knowledge) but differed in the delivery method and level of interactivity. Details of each tour are presented in Section 2.2. In a driving simulator experiment, we compared the effectiveness of the familiarization tours on automation trust and driving performance in critical and non-critical transition tasks.

## 2. Methods

*2.1. Participants*

Ninety-seven volunteers were recruited from three colleges with approximately 5,600 students and 450 employees using flyers and student social groups on Facebook. After screening, 84 of those volunteers were eligible to participate in the experiment and were randomly assigned to one of the four groups (no familiarization, video, low-fidelity VR, and high-fidelity VR), with 21 participants in each group. Female participants comprised 45.2% of the participants, and the mean age was 24.18 years (SD = 8.3). As recruiting criteria, all the participants had no or little familiarity with highly automated cars and had no or little experience with VR. All participants completed the experiment, but data on a few transition tasks missed for three participants.

*2.2. Familiarization tours*

As illustrated in Table 1, participants in the video group received a tour using a monitor. In the low-fidelity VR group, the familiarization tour was delivered through an HMD VR where information was presented in an Oculus Go headset, and participants were able to interact with the virtual environment using a handheld game controller. This controller allowed them to manually drive the car and engage or disengage the automated mode using a reserved button. Participants in the high-fidelity VR tour instead interacted with the virtual environment using a steering wheel and pedal system. To control the level of fidelity and immersion in VR groups, participants used the same VR headsets and received identical materials with the same level of visual realism and spatial layouts. Though the training materials and scenarios and transition tasks were the same in all three tours (Figure 1), participants in the video group read the explanation and watched the videos displayed on the monitor, and participants in VR groups received this information in an interactive VR environment.

The materials in the video tour were made using Microsoft PowerPoint and Animaker. The VR tour was created using Unity v5 (Unity Technologies) and was integrated into Oculus headsets using Oculus VR SDK. Oculus VR SDK for Unity is a package provided by Oculus to integrate VR environments built in Unity with Oculus VR headsets. We used their rendering features, interface for controlling VR camera behavior, a first-person control prefab, and the unified input API for controllers. A 3-D model of Tesla S P100D was used, and the models of road environments (e.g., urban area, highways, rural road, road objects) were imported from the EasyRoad package and edited and assessed for accuracy and visibility in the virtual environment. The Oculus Go device was run on a machine with a CORE i7 CPU @ 3.30GHz, 8GB RAM with OS/Windows 10 x64 operating system.

*Table 1.* Specifics of the familiarization tours (video, low-fidelity VR, and high-fidelity VR)

| Tour | Video | Low-fidelity interaction VR | High-fidelity interaction VR |
|---|---|---|---|
| | 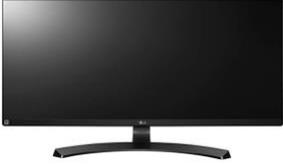 | 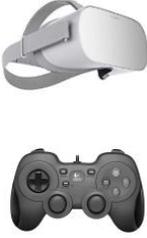 | 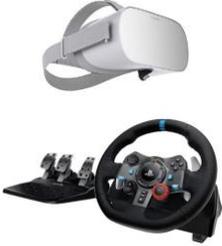 |
| Display | Flat screen | Oculus Go, HMD | Oculus Go, HMD |
| Screen resolution | 2560 x 1080 34"1080p | 2560×1440 5.5" (538ppi) fast-switching LCD with standard 60Hz refresh, "overclocked" 72Hz refresh | 2560×1440 5.5" (538ppi) fast-switching LCD with standard 60Hz refresh, "overclocked" 72Hz refresh |
| Input device movement | No input, no interaction | Logitech Dual Action GamePad | Logitech Driving Force G29 Racing Wheel with Pedals |
| Training paradigm | Automation, driver, task knowledge (video + texts explanation on a monitor) | Automation, driver, task knowledge and practice (video + texts explanation + interactive practices) | Automation, driver, task knowledge and practice (video + text explanation + interactive practices) |
| Pre-trip tour process | • 3 slides (text + video) for automation system features (ACC and LKA)<br>• 2 slides (text + video) for physical equipment (camera, radar, ultrasonic sensor, GPS sensor)<br>• 2 slides (text + video) for driver's knowledge<br>• 30-minute video about three types of transitions and use cases (3 use cases for critical takeover, 2 use cases for a non-critical takeover, and 2 use cases for handback tasks) | • 7 minutes (on average) interactive explanation (text + video) for automation system features (ACC, LKA)<br>• 10-minute (on average) walkaround the car to observe physical equipment (camera, radar, ultrasonic sensor, GPS sensor)<br>• 10 minutes (text + video) for driver's knowledge<br>• 30-minute (on average) practices and explanation of three types of transitions and use cases (3 use cases for critical takeover, 2 use cases for non-critical takeover, and 2 use cases for handback tasks) using a hand-held game controller | • 7 minutes (on average) interactive explanation text + video) for automation system features (ACC, LKA)<br>• 10-minute (on average) walkaround the car to observe physical equipment (camera, radar, ultrasonic sensor, GPS sensor)<br>• 10 minutes (text + video) for driver's knowledge<br>• 30-minute (on average) practices and explanation of three types of transitions and use cases (3 use cases for critical takeover, 2 use cases for non-critical takeover, and 2 use cases for handback tasks) using a steering wheel and pedals |

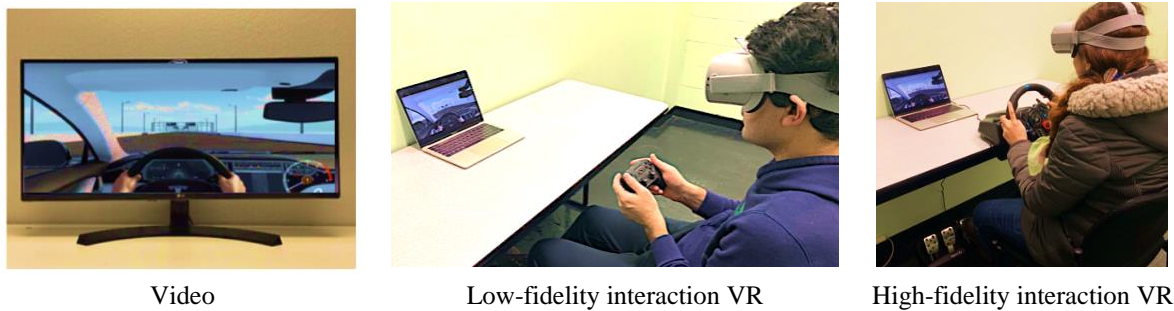

| Video | Low-fidelity interaction VR | High-fidelity interaction VR |

*Figure 2.* The familiarization tours: video, low-fidelity VR, and high-fidelity VR

## 2.3. Driving simulator

This experiment was performed using a fixed-based simulator, which was designed in the Unity game engine and operated on Dell Optiplex 7010 (Intel Quad-Core i7-3470 3.2GHz, 16GB RAM) workstation running Windows 10. Three widescreen displays showed the visual simulation imagery, rendered at 60 Hz. The simulator was able to provide two driving modes: level 3, or conditionally automated driving, and manual driving. Based on the features outlined for level 3 automation in SAE J2016, the automated mode supported simultaneous longitudinal and lateral control and enabled the car to detect objects, execute independent lane-change maneuvers, and request a takeover when necessary. Participants were able to engage and disengage automaton by pressing the same button located on the right side of the steering wheel. Disengagement was also possible through pressing the brake (> 10% of braking length) or turning the steering wheel more than (> 7 degrees). A takeover request (TOR) was also issued in an auditory format with a high-pitched tone ("In the next ten seconds, take over the control") through two speakers placed near the participant. Ten seconds is suggested as an appropriate length of time in which drivers can safely switch to manual mode [46]. All TORs had identical features with regard to format and wording, with the only difference being the respective cause and action. Table 2 depicts the visual symbols of HMIs used to display the status of automation mode. These visual symbols were displayed on the right-bottom corner of the middle display.

*Table 2.* Visual symbols of HMI showing the status of automation mode

| Automation is running | Automation is ready to use | Takeover is required (TOR is issued) | Automation is disengaged |
|---|---|---|---|
| 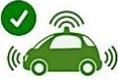 | 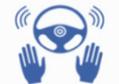 | 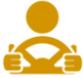 | 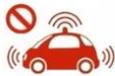 |

## 2.4. Design of experiment

This study employed a mixed between-within design, with the type of familiarization (no familiarization, video, low-fidelity VR, high-fidelity VR) as the between-subjects factor, as well as transition scenarios

(critical takeover, non-critical takeover, and handback) and time of measurement as within-subjects variables. Participants in the no familiarization, or control group (n = 21), were introduced to the drive test without receiving any form of a familiarization tour. Before the drive tests, participants in the video group (n = 21), low-fidelity VR (n = 21), and high-fidelity VR (n = 21) were familiarized with the highly automated driving using a screen, a low-fidelity interaction, and a high-fidelity interaction VR, respectively. In the drive tests, participants in all groups were exposed to the same simulator, driving environment, and scenarios. Moreover, all participants were assigned to engage in an identical secondary task (watching video). They experienced three types of transitions – critical takeover, non-critical takeover, and handback (Table 3) – three times in different blocks. The sequence of these scenarios in each block was randomly specified to avoid learning effects.

*Table 3.* Three types of transition: critical takeover, non-critical takeover, and handback

| Critical TOR scenario | Non-critical TOR scenario | Handback scenario |
|---|---|---|
| 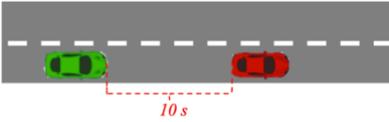 | 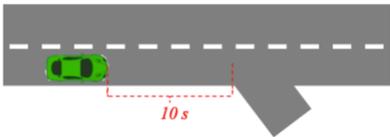 | 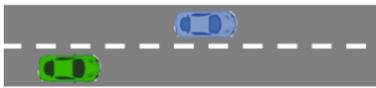 |
| The red car has blocked the right lane, and the subject car (green) issues a TOR with a time budget of 10s, stating that the lane is blocked and takeover is necessary. | The subject car (green) issues a TOR with a time budget of 10s, stating that a scheduled exit is ahead. | The subject car issues a handback message stating that automation is available. |

### 2.5. Procedure

Each participant was individually welcomed into the experiment room and received a written informed consent with a notice that they could end the experiment at any time without providing a reason and without negative consequences. After having signed the informed consent documents, each participant read the instructions to complete the experimental task. The participants were then asked to complete the demographic questionnaire. After completing the questionnaires, participants briefed on the simulator. In the next step, participants who were assigned to take the video tour received a one-hour tour involving text and video covering the previously described content (Figure 1 and Table 1) to familiarize themselves with automated driving. In VR groups, participants received the tour for approximately one hour, then moved to the drive test. Before starting the drive test, all participants were asked to take the knowledge tests; further, participants in training groups were also asked to complete a set of questionnaires: presence and VR sickness questionnaires. Before beginning the drive test, participants were given the option to take a 30-minute break. The drive tests contained three blocks; each one covered three types of transitions (critical takeover, non-critical takeover, and handback). In each block, participants were presented with an urban section that included intersections, pedestrians, and a single and dual carriageway with heavy traffic. They were then directed to a highway section, representing three-lane roads with ramps and traffic signs. At the end of each block, participants were administered the trust questionnaire (Figure 3). When the automated mode was active, participants were required to watch a video of the Our Planet series on Netflix, which was displayed on a tablet mounted to the dashboard. After finishing the experiment, the participants were debriefed and thanked for their time.

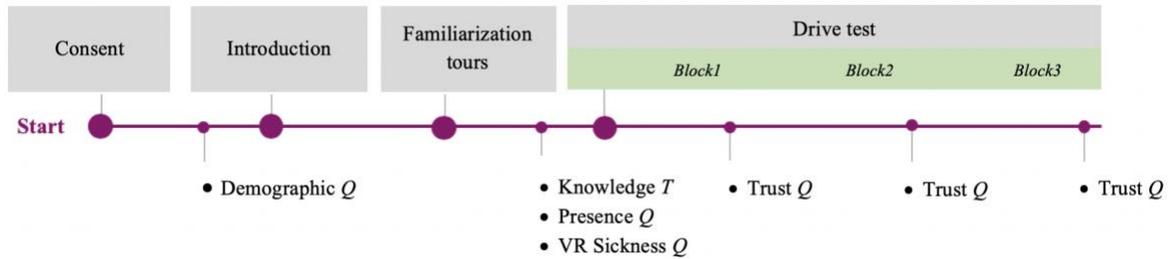

Figure 3. Experiment procedure; Note 1: Participants in the control group skipped the familiarization tour and its questionnaires; Note 2: Q denotes a questionnaire and a T denotes a test

*2.6. Measures*

Two sets of variables associated with learning outcomes and learning experiences were measured to answer the research questions.

*2.6.1 Learning outcomes*

*Takeover time (TOT)* in critical and non-critical scenarios: Takeover time was measured from the moment a TOR was issued until the participant took control of the car using the steering wheel, acceleration pedal, or brake pedal.
*Handback time (HBT)*: Handback time measured the time it took for a participant to relinquish control to the automation when he or she was informed that the automation was available to control the car.
*Takeover quality* was evaluated by *minimum time to collision (TTC)* and *SD of lane position (SDLP)* during one mile of the manual driving period after the TOR.
*Trust*: A 12-item questionnaire proposed by [47] was used to measure self-reported trust. Mean self-reported trust was calculated for each participant by averaging individual values.
*Verbal knowledge*: A questionnaire was used to measure participants' verbal knowledge of highly automated driving.

*2.6.2 Learning experience*

*Presence*: The Igroup Presence Questionnaire (IPQ) [48] was used to measure self-reported presence (sense of "being there") in four areas – sense of being there, spatial presence, involvement, and experienced realism – on a 7-point Likert scale, ranging from 0 to 6 for each item.

*VR sickness*: The general items of the Simulator Sickness Questionnaire (SSQ) [49] were used to measure the level of adverse health effects of the VR tour. The general items of this tool consisted of nine questions with a 4-point scale ("none", "slight", "moderate", or "severe").

*2.7. Hypotheses*

H1: a) Takeover time in both critical and non-critical scenarios will be shorter in familiarization groups, particularly in VR tours. b) Throughout the experiment, takeover time will decrease and finally stabilize.

H2: a) Takeover quality is more likely to be higher (longer minimum TTC and smaller SDLP) among educated participants. b) In particular, increasing the level of interaction fidelity in VR will improve takeover quality.

H3: a) Handback time will decrease as a result of the familiarization tours. b) This improvement will be higher in interactive familiarization tours. c) Over the course of the experiment, handback time will decrease and finally stabilize.

H4: a) Participants in training groups are more likely to report higher automation trust after experiencing failures compared to the control group. b) Self-reported automation trust will increase and finally stabilize after several exposures to TORs.

H5: a) Verbal knowledge score will be higher when participants receive a familiarization tour. b) VR tours will help participants attain more knowledge.

H6: High-fidelity VR increases the level of presence.

## 3. Results

### 3.1. Reaction performance

*TOT of critical scenarios*: An ANOVA test revealed a significant effect of the type of familiarization tour on TOT of critical scenarios at the first failure, $F(3, 68) = 5.29, p < 0.01, M\ control = 4.63\ s\ (2.23), M\ video = 3.72\ s\ (1.35), M\ low\text{-}fidelity\ VR = 2.94\ s\ (0.53), and\ M\ high\text{-}fidelity\ VR = 2.83\ s\ (0.95)$. According to the post-hoc test, aside from the comparison between low-fidelity VR and high-fidelity VR, the rest of the comparisons were significant. No statistically meaningful difference was observed in TOT between tour groups at the second and third times of measurement. A significant interaction between familiarization tour and time of measurement ($F[4, 205] = 5.08, p < 0.05$) effect was observed in TOT of critical scenarios. As demonstrated in Figure 4, TOT of the control and video groups was significantly shorter at the third failure than at the first and second. The difference between the second and third failures was not significant in any group.

*TOT non-critical scenarios*: TOT in non-critical scenarios was also meaningfully affected by the type of familiarization tour at the first failure, $F(3, 68) = 4.89, p < 0.05$ (Figure 5). Compared to the control group ($M = 4.17\ s, SD = 2.13$), participants in the high-fidelity VR group demonstrated significantly shorter TOT ($M = 2.97\ s, SD = 1.03$). No difference was found in TOT between groups at the second and third failures. Time of measurement had a main effect on the control group such that TOT significantly decreased from the first failure ($M = 4.17\ s, SD = 2.13$) to the third ($M = 2.81\ s, SD = 0.94$). Time of measurement had no significant impact on the TOT of participants who received a familiarization tour.

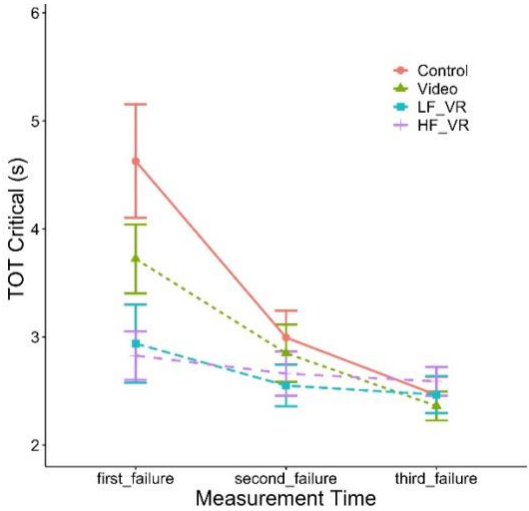 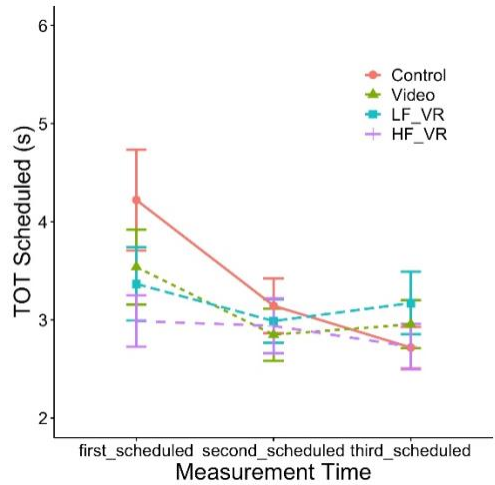

*Figure 4.* Average TOT in critical scenarios (± SE) in different times of measurement for four familiarization tours (control, video, low fidelity [LF] VR, and high-fidelity [HF] VR)

*Figure 5.* Average TOT in non-critical scenarios (± SE) in different times of measurement for four familiarization tours (control, video, low-fidelity [LF] VR, and high-fidelity [HF] VR)

*HBT*: The ANOVA test confirmed a significant effect of familiarization tour ($F [3, 65] = 8.46, p = 0.001$) at the first block, as well as time of measurement ($F [2, 51] = 4.33, p = 0.02$) for the control group (Figure 6). Pairwise comparisons with Bonferroni-corrected adjustments revealed a meaningfully longer takeover time only in the control group at the first block. Moreover, pairwise comparisons revealed a significant difference between the first handback scenarios and the second ($p < 0.05$) and third ($p < 0.01$), demonstrating a longer mean takeover time at the first handback events. No significant effect was observed either between the second and third exposure to handback requests or between tour groups (video, low-fidelity, and high-fidelity VRs) at any time of measurement.

### 3.2. Takeover quality

*TTC*: Type of familiarization tour had no significant impact on longitudinal vehicle control in critical takeover scenarios, $F (3, 209) = 1.82, p = 0.103$; however, when the measurement time and type of tour were taken into account, a significant interaction effect was found in TTC. Compared to the first failure, TTC of participants in the control and video groups significantly increased at the third failure (*control: F (2, 56) = 5.72, p < 0.01; video: F (2, 56) = 4.85, p < 0.05*). Other contrasts between the second and third failures across familiarization groups did not reveal any significant relationship.

*SDLP*: Type of tour had a significant main effect on lateral vehicle control within one mile after TOR (*SDLP: F [3, 210] = 6.71, p < 0.01*). A Tukey post-hoc test revealed that participants in the high-fidelity VR group showed significantly smaller SDLP (*M = 38 cm, SD = 13.1, p < 0.01*) compared to the other three groups. This effect, however, disappeared after the second critical takeover scenario. SDLP was affected by the time of measurement, $F (2, 210) = 5.03, p < 0.05$. All participants at the last measurement demonstrated a smaller deviation in lateral position; however, this change was not significant in the high-fidelity VR group (p = 0.061).

### 3.3. Automation trust

A Kruskal-Wallis test revealed a main effect of familiarization tours ($\chi2(3) = 14.21, p < 0.01$) and time of measurement ($\chi2(2) = 11.499, p < 0.01$) on self-report automation trust (Figure 7). While no significant change was observed in participants in VR groups across all three blocks, trust rating among participants in the control and video groups significantly increased from block 1 (*M control = 3.73[1.34], M video = 4.13[1.33]*) to block 3 (*M control = 5.18[1.20], M video= 5.39 [1.46]*). Another noteworthy finding is that, at the end of the third block, no significant difference was observed between distinct groups. Regarding the effects of the block, the paired exact Wilcoxon test revealed that there were significant differences only between the first and third blocks ($Z = 3.09, p < 0.01$); $p > 0.09$ for all other differences. Trust rating at the end of the third block was significantly higher than in the first block in the control and video groups. A post-hoc bilateral Mann–Whitney U test also only revealed a meaningful difference when comparing the control group with low-fidelity VR ($p < 0.05$) and high-fidelity VR ($p < 0.05$).

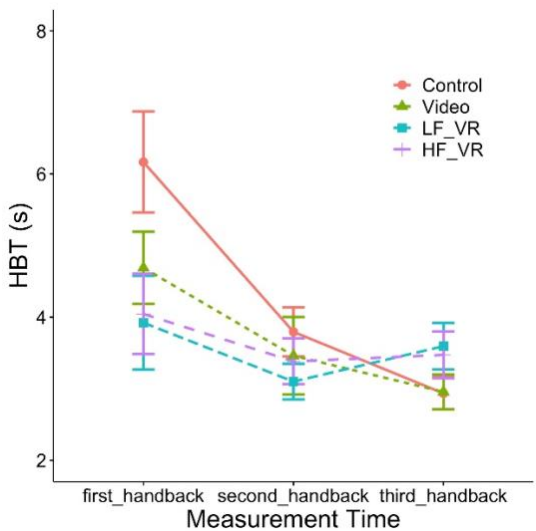
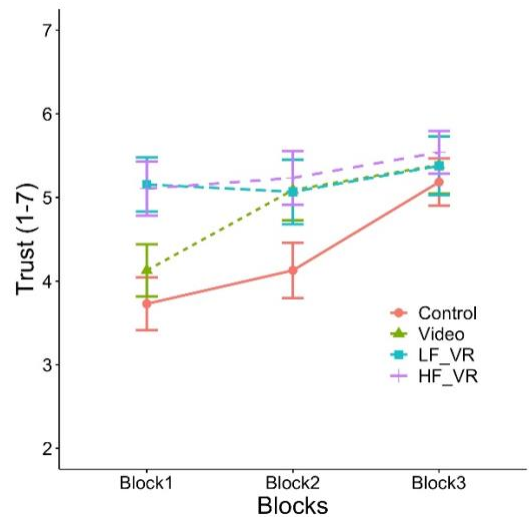

Figure 6. Average HBT (± SE) in different times of measurement for the four familiarization tour groups (control, video, low-fidelity [LF] VR, and high-fidelity [HF] VR).

Figure 7. Average trust (± SE) in different times of measurement for the four familiarization tour groups (control, video, low-fidelity [LF] VR, and high-fidelity [HF] VR).

### 3.4. Knowledge test

After completing the tours and before the drive test, participants in all groups were given a 15-question test measuring their level of knowledge. The control group received a mean score of 31.7% (9.39), video group 46.1% (6.1), low-fidelity VR 63% (14.8), and high-fidelity VR 65.8% (12.4). The difference in test scores between the control and participants who received one of the familiarization tours was found to be significant ($p < 0.01$). Compared to the video group, participants in VR groups also achieved a higher knowledge score ($p < 0.05$); however, there was no main effect of the type of VR (interaction fidelity) tour on knowledge test scores ($p = 0.21$).

### 3.5. Presence

As shown in Figure 8, the level of presence (total score) was significantly higher in the high-fidelity VR ($M = 3.59, SD = 0.58$) and low-fidelity VR ($M = 2.91, SD = 0.86$) than in the video tour ($M = 2.05, SD = 0.47$) ($\chi_2 (2) = 94.5, p <0.01$). A Mann-Whitney post-hoc test additionally revealed that all comparisons were statistically significant ($p < 0.01$). Comparing VR tours in each item indicated that participants in the high-fidelity VR group experienced significantly higher involvement ($p < 0.05$) and spatial immersion ($p < 0.05$). No other significant difference was observed in the other two items of IPQ between VR tours.

As depicted in Table 4., both virtual reality tours resulted in similar symptom experiences during the training; however, the severity of dizziness ($p < 0.05$) and disorientation ($p < 0.01$) were significantly higher among participants in the high-fidelity VR group. In general, compared to the low-fidelity VR group ($M\ total\ score = 5.3, SD = 1.31$), the total score was higher in the high-fidelity VR group ($M\ total\ score = 6.81, SD = 2.06$); however, it was not statistically significant ($p = 0.13$).

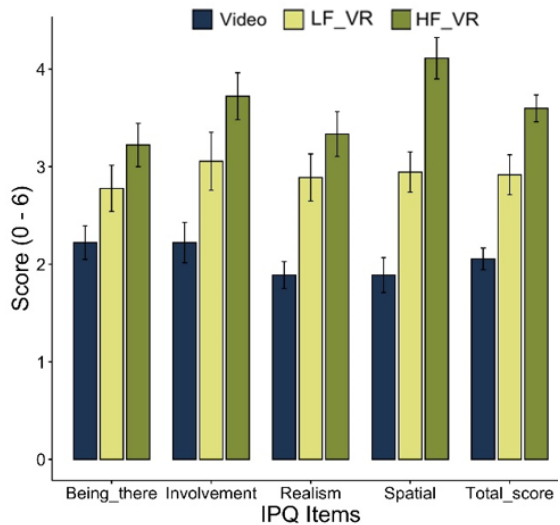

*Figure 8.* Average score of IPQ (± SE) for familiarization tours (video, low-fidelity VR, and high-fidelity VR)

*Table 4.* Percentage of participants reporting severe or moderate levels for each item in SSQ

| Symptoms | Low-fidelity VR (%) | High-fidelity VR (%) | *p* value |
| --- | --- | --- | --- |
| General discomfort | 8 | 5 | - |
| Fatigue | 10 | 7 | - |
| Boredom | 0 | 0 | - |
| Drowsiness | 8 | 0 | - |
| Headache | 12 | 15 | - |
| Dizziness | 23 | 38 | * |
| Nausea | 17 | 21 | - |
| Disorientation | 19 | 43 | ** |

Mann–Whitney U test: *$p < 0.05$, **$p < 0.01$

# 4. Discussion

Through a driving simulator experiment with 84 participants, the current study investigated how different types of familiarization tour (video, low-fidelity VR, and high-fidelity VR) affect trust, transition time and quality in takeover (critical and non-critical) and handback situations.

*4.1. Impact of familiarization tours on takeover time*

The results partially support our hypothesis regarding TOT in both critical and non-critical scenarios (H1a), in that all three familiarization tours significantly improved TOT. However, this effect was only significant at the first critical and non-critical TOR events; takeover performance in subsequent situations converged to the high-fidelity VR group (Figures 3 and 4). This is line with [50], who showed that prior familiarization with highly automated driving had more positive effects on TOT at the first takeover situation than the subsequent events. The results also indicated a positive impact of the use of VR and interactive practices on takeover time in both critical and non-critical situations. Similarly, [23, 35] revealed a faster takeover time as a result of training with VR. It seems that VR with interactive takeover tasks is an effective delivery method to educate first-time users to respond faster upon receiving TORs.

Our findings also indicate that participants in the video tour performed faster than those in the control group. This may imply that explaining the automation features within the use cases helps first-time users build more accurate mental models of how the automation works and consequently react faster to TORs. However, the results related to the effectiveness of non-interactive training, for example, text-based or video-based presentation of the automation features, such as what manufactures provide in owner manuals, is controversial. While written description of automation features, as demonstrated by [50], or showing takeover use cases, such as in [24], have been found to improve performance, [51] found that the written system information did not support the participants in building sufficient mental models of the system and making correct decisions on whether takeover was necessary. These inconsistencies stem from several sources of variation observed in training protocols, methodology, design of experiment, drive tests, and metrics. Training protocols, for example, share only a few common elements in feature explanation, use cases, or delivery methods, which makes it difficult to reach conclusive findings.

Congruent with [50], time of measurement was also an influential factor in TOT of critical and non-critical takeover situations (H1b). As illustrated in Figures 3 and 4, takeover time tended to decrease over the driving sessions and finally stabilize. [52] suggests that the driver's behavioral measures in transition tasks follow the power law of practice, such that there is an increase in takeover performance at the beginning of the experience and stabilization after about three trials. However, this pattern was only observed in the control and video groups in the critical takeover scenario (Figure 4) and the control group in the non-critical takeover task (Figure 5). One potential explanation is that participants in VR groups were exposed to several interactive takeover practices before the drive test. They may have reaped the benefits of behavioral improvement offered in training sessions and were close to the stabilization phase.

*4.2. Impact of familiarization tours on takeover quality*

In contrast to part of our hypothesis (H3 a and b), there was no main effect of familiarization tours on minimum TTC, though a significant interaction effect between the type of tour and time of measurement was observed in minimum TTC data. This finding is consistent with that of [50], who reported a similar interaction effect between the time of measurement and type of training. According to their results, the minimum TTC of no-familiarization and experience groups was significantly lower at the first than at the second time of measurement; compared to prior experience, prior description of transition tasks had a more positive impact on takeover performance in first failures. Our findings, however, are not fully consistent with their results. Participants in the video group who had just received a description of the automation system did not perform better than the control group in the first failure, and instead experienced a meaningful improvement in the second and third failures. Differences in descriptions provided in these studies could be a source of this conflict.

The results partially support our hypothesis regarding the effectiveness of the familiarization tour on SDLP (H3 a). Only participants who practiced takeover situations using VR with high-fidelity interaction exhibited smaller SDLP at the first failure. It seems that improving the fidelity of interaction in a VR tour could help novice drivers complete the takeover process smoother than those who did not practice the tasks with steering wheel and pedals before facing the first failure. Likewise, [53] found higher stability of control recovery in semi-automated driving as a result of improving interaction fidelity from two handheld controllers to a steering wheel and pedals. There are two potential explanations for this finding. First, improving interaction fidelity [54, 55] and controller naturalness [56, 57] increases the level of immersion in virtual environments, resulting in a higher sense of spatial presence (Figure 8). Being more immersed in the learning environment may help the learner achieve higher learning transferability in sensory-motoric levels. Second, when the movements are similar to reality, more opportunities are provided to transfer motor skills [58] by developing operational behavior in skill-based level performance, which requires a high level of sensory-motoric coordination. The similarity of body movements (steering wheel and pedals) in the high-fidelity interaction VR tour with drive tasks refines the takeover quality and leads to less erratic deviation from the center of the lane after regaining manual control.

In general, the findings regarding takeover performance demonstrate the positive impact of exposing the driver to interactive practices, especially when they are presented in high-fidelity interaction VR. This finding is supported by earlier research in automated driving and other domains indicating the influential role of practice trials on facilitating the learning process and developing procedural skills. Though mastering the sensory-motoric skills required for a smooth and efficient takeover performance demands an extensive amount of repetition, engaging drivers in a few simulated takeover events prior to the first trip may resemble the associative stage [17], where strategies are formulated and drivers are able to build extensive mental models of the automation. In these intermediate stages of skill acquisition, the resources needed for cognitive processes decrease, and the quality (speed and accuracy) of takeover performance increases. Receiving extensive and continuous exposure to a task is expected to refine skills learned in the previous phase and make behavior more automatic and unconscious with further use (autonomous or procedural stage). In highly automated driving, however, considering the low frequency of critical takeover situations, it is less likely that one can quickly move to the autonomous stage and develop psychometric skills at the autonomic level. Beyond the prior familiarization tour, periodic training in forms of in-car tutoring or driving schools could refresh stored rules and activate knowledge of sequences of interrelated actions and reactions that are expected to occur in critical takeover situations.

*4.3. Impact of familiarization tours on handback time*

Familiarization tours were found to contribute to decreasing handback time (H4 a). Educated participants relinquished control to the automation in a shorter time; however, like the patterns observed in TOT data, this was only found at the first handback request (Figure 6). The transition from manual to automated modes has not been the topic of many studies thus far and seems less complex, though it raises more ethical and legal issues, particularly in the system-initiated handback tasks [59]. In the current study, however, we only considered a human-initiated handback situation. In these scenarios, participants were told to delegate control to the automation by pressing a button on the steering wheel as soon as they received handback requests as auditory and visual feedback (Table 2, "Automation is ready to use"). Such a task may not realistically simulate some of the common handback activities on real roads, for example, when the driver is not forced to relinquish manual control, and the transition to the automated mode is more of a top-down decision-making process, rather than bottom-up processing.

*Impact of familiarization tours on automation trust*

Compared to the control and video groups, participants who received VR-based and interactive practices reported higher automation trust after the first failure; moreover, their level of trust did not change significantly from the first failure to the last failure, indicating more consistent and calibrated trust. In support of this finding, [60] demonstrated the influential impact of practice trials to mitigate the adverse effects of overtrust. Later, [25] additionally observed more frequent glances to the road – indicating distrustful behavior – among participants who only received simple training (text and video explanation). It seems that when drivers clearly understand how the automation works, they are more likely to build a more calibrated level of automation trust. As suggested by [61], our findings confirm that when training provides information regarding the purpose, process, and performance of automation in interactive contexts, the appropriateness of trust increases.

In line with previous studies [62], our results revealed a significant main effect of time of measurement on self-reported automation trust rating. Automation trust increased during the driving session. However, this pattern was significantly affected by familiarization group, such that only participants in the control and video groups exhibited this pattern, while participants in VR groups reported a relatively similar level of trust in all three measurements. Interaction of time of measurement and level of familiarity with the automation features was also reported [63]. In lower levels of automation, [8] also found the trust to ACC tended to stabilize and calibrate after several experiences with the system (five trips). Observing the system behavior [61] and accumulating the knowledge of underlying processes from interactions [64] advance the human mental models of automation and establish more calibrated trust.

*4.4. Impact of familiarization tours on knowledge and presence*

Regardless of the type of familiarization, participants who received a tour gained knowledge of the automation. However, knowledge acquisition was meaningfully higher when the content was delivered through VR. This finding is congruent with a large body of existing evidence suggesting VR-based learning as a more effective method of knowledge acquisition and retention compared to traditional methods [65]. Higher affordances of VR in presenting information more intuitively leads higher immersion and may increase the trainee's motivation to learn. Moreover, our findings indicate that participants who received the VR tour with higher-level interaction fidelity (high-fidelity VR) did not achieve higher knowledge scores. It seems that, although delivering content using VR can interactively improve knowledge acquisition, interaction fidelity does not matter in this case.

Despite the results of previous studies indicating a positive association between the level of immersion and knowledge acquisition [66-68], our findings failed to demonstrate a fully consistent positive relationship between level of immersion and knowledge score. Although improving the level of immersion from the video tour to low-fidelity VR led a higher sense of presence and a consequently higher knowledge score, increasing the level of immersion in high-fidelity VR only heightened the sense of presence and did not lead to more learning. Recent findings in immersive VR learning [69-72] suggest that increasing the level of immersion (and presence) in virtual environments (VEs) does not necessarily increase knowledge acquisition; further, even in some highly immersive VEs, due to redundancy of irrelevant information, learning outcomes are fewer than in VEs with lower immersion. One of the shortcomings in immersive learning is that there are still insufficient and inconclusive studies on which technical aspects of a VR environment, and to what extent, contribute to presence to support knowledge and skill acquisition.

## 5. Limitations and future directions

The results of this study need to be interpreted in consideration of several limitations. As discussed by [62] and [7], when participants expect and are aware of failures in a short period, the level of trust does not decrease. Even when we randomized the events and distributed them at different time intervals, it was evident that they expected some failures. In reality, however, failures do not occur with such frequency. Further simulated and real-road studies are necessary to explore this association in longer driving sessions with less frequent failures. Moreover, type of secondary tasks has been shown as an important factor in level of disengagement from driving loop [73]. Future studies are required to the effectiveness of training programs while participants are engaged in different secondary tasks.

We adopted the training requirements outlined by [11], which focused on partially automated driving. Since the driver's role in highly automated driving is significantly different from in a lower level of automation, future research should explore a comprehensive training protocol for highly automated driving. Such a protocol not only needs to focus on materials and appropriate delivery methods; it also needs to encompass testing scenarios and appropriate metrics to measure and compare learning outcomes systematically.

One may expect that, due to the higher urgency of critical scenarios, TOT in such events would be shorter than in the non-critical TOR situations; however, our findings did not reveal such a meaningful difference. We did not measure the subjective criticality of the situations and are not able to confirm whether participants perceived a comparable level of criticality for non-critical and critical events. Future studies should consider subjective criticality (e.g., [50, 74]) to compare the driver's perception of different takeover scenarios and evaluate the effectiveness of training on driver risk perception.

We recruited participants who were not familiar with highly automated driving. In reality, however, people may have acquired prior knowledge from friends, family, and other information sources (word of mouth) and built initial trust and opinions about the system. [61, 75] argue that rumors and gossip can have detrimental effects on trust regarding system capabilities; therefore, these factors should be considered in training and retraining operators. Moreover, perceived usefulness was reported as a strong predictor of user intention in highly automated driving [76, 77]. Future research should explore the effects of training and misinformation on technology acceptance and driving performance in highly automated driving.

Moreover, the performance of a driver with a high level of experience with manual driving (but who is a novice in automated driving) differs from that of a driver with no or little manual driving experience. An experienced driver is equipped with an extensive mental model of various traffic environments, the dynamicity of roads, and other vehicles' behavior (rules and knowledge), which helps him or her maintain a better SA [78]. Moreover, a novice driver does not enjoy the level of proficiency that an experienced driver has with sensory-motoric behavior, such as smooth braking and steering in a lane change task. Previous studies also have shown that users' socio-demographics, cultural, and experience affect their behavioral pattern in using technologies [79, 80]. Future studies should consider these potentially mediating factors in drivers behavior and their experiences in self-driving vehicles.

## 6. Conclusion

Although numerous works have studied highly automated driving, only a few studies have been carried out on training and educating first-time users of highly automated cars. In this study, three types of familiarization tour (video, low-fidelity VR, and high-fidelity VR) were examined in a driving simulator experiment. Our findings contribute to the existing knowledge on educating first-time users of highly automated vehicles in several ways. First, familiarizing first-time users with highly automated driving, particularly when interactive practice trials are provided, positively affected trust and transition performance. Second, VR tours, especially when presented in higher interaction fidelity, had a positive impact on takeover performance and trust. Third, regardless of the type of transition, participants' performance at the first time measurement significantly differed from the subsequent situations, indicating higher costs associated with first time takeover for uneducated and less-educated users. This necessitates that policy makers focus on assembling regulations and standards related to training curriculum and issuing licenses for automated driving. Based on these findings, further research is necessary to examine how these results can be replicated in real driving conditions.

## 7. Acknowledgment

The authors would like to thank all people that have been involved in the designing VR programs and implementation of the study. In particular, we would like to thank Amanda Seccia, Dave Song, and Hamid Salmani for their contributions in the study. This project did not receive any specific grant from funding agencies in the public, commercial, or non-profit sectors.